\documentclass[aps,prl,twocolumn,showpacs,superscriptaddress]{revtex4}
\usepackage{graphicx}

\begin{document}

\title{Electronic self-organization in the single-layer manganite
$\rm Pr_{\it 1-x}Ca_{\it 1+x}MnO_4$}

\author{F. Ye}
\email{yef1@ornl.gov}
\affiliation{Neutron Scattering Science Division,
Oak Ridge National Laboratory, Oak Ridge, Tennessee 37831-6393,
USA}
\author{Songxue Chi}
\affiliation{Department of Physics and Astronomy,
The University of Tennessee, Knoxville, Tennessee 37996-1200, USA}
\author{J. A. Fernandez-Baca}
\affiliation{Neutron Scattering Science Division,
Oak Ridge National Laboratory, Oak Ridge, Tennessee 37831-6393,
USA}
\affiliation{Department of Physics and Astronomy,
The University of Tennessee, Knoxville, Tennessee 37996-1200, USA}
\author{A. Moreo}
\affiliation{Department of Physics and Astronomy,
The University of Tennessee, Knoxville, Tennessee 37996-1200, USA}
\affiliation{Materials Science and Technology Division,
Oak Ridge National Laboratory, Oak Ridge, Tennessee 37831-6393,
USA}
\author{E. Dagotto}
\affiliation{Department of Physics and Astronomy,
The University of Tennessee, Knoxville, Tennessee 37996-1200, USA}
\affiliation{Materials Science and Technology Division,
Oak Ridge National Laboratory, Oak Ridge, Tennessee 37831-6393,
USA}
\author{J. W.~Lynn}
\affiliation{NIST Center for Neutron Research, Gaithersburg,
Maryland, 20899, USA}
\author{R. Mathieu}
\affiliation{ERATO Spin Superstructure Project and Multiferroics
Project, JST, Tokyo 113-8656, Japan}
\author{Y. Kaneko}
\affiliation{ERATO Spin Superstructure Project and Multiferroics
Project, JST, Tokyo 113-8656, Japan}
\author{Y. Tokura}
\affiliation{ERATO Spin Superstructure Project and Multiferroics
Project, JST, Tokyo 113-8656, Japan}
\affiliation{Cross-Correlated Materials Research Group (CMRG), RIKEN
Advanced Science Institute, Wako 351-0198, Japan}
\affiliation{Department of Applied Physics, University of Tokyo,
Tokyo 113-8656, Japan}
\author{Pengcheng~Dai}
\affiliation{Department of Physics and Astronomy,
The University of Tennessee, Knoxville, Tennessee 37996-1200, USA}
\affiliation{Neutron Scattering Science Division,
Oak Ridge National Laboratory, Oak Ridge, Tennessee 37831-6393,
USA}
\date{\today}

\begin{abstract}
We use neutron scattering to investigate the doping evolution of the
magnetic correlations in the single-layer manganite $\rm
Pr_{\it 1-x}Ca_{\it 1+x}MnO_4$, away from the $x=0.5$ composition where the
CE-type commensurate antiferromagnetic (AF) structure is stable.  We
find that short-range incommensurate spin correlations develop as
the system is electron doped ($x<0.5$), which coexist with the
CE-type AF order.  This suggests that electron doping in this system
induces an inhomogeneous electronic self-organization, where
commensurate AF patches with $x=0.5$ are separated by electron-rich
domain walls with short range magnetic correlations.  This behavior
is strikingly different than for the perovksite $\rm Pr_{\it
1-x}Ca_{\it x}MnO_3$, where the long-range CE-type commensurate AF
structure is stable over a wide range of electron or hole doping
around $x=0.5$.
\end{abstract}

\pacs{75.47.Lx, 75.30.Kz, 25.40.Dn}

\maketitle

Understanding how electrons are distributed in narrow bandwidth
manganese oxides remains one of most intriguing unresolved problems
in the physics of colossal magnetoresistance (CMR) manganites
\cite{dagotto01}.  In the perovksite $\rm R_{\it 1-x}Ca_xMnO_3$
($R=$ rare earth ions) and single-layer $\rm R_{\it 1-x}Ca_{\it
1+x}MnO_4$ manganites (the $n=\infty$ and $n=1$ end members of the
Ruddlesden-Popper series $\rm R_{\it n(1-x)}Ca_{\it nx+1}Mn_{\it
n}O_{\it 3n+1}$ manganese oxides), the orbitals and magnetic spins
of the Mn ions order at low temperature at the doping level $x=0.5$,
and form a checkerboard-like pattern, resulting in a commensurate
(CM) antiferromagnetic (AF) CE-structure
\cite{wollan55,sternlieb96,larochelle01}. 
Our research here clarifies what occurs in the manganites near
$x=0.5$ when their structures evolve from the three-dimensional (3D)
perovskite to the single layer structure which is the most
two-dimensional (2D) member of the Ruddlesden-Popper series. Are
novel charge-ordered states formed, or do the excess of
electrons/holes become randomly distributed and localized, or does a
self-organization occur involving mixed-phase tendencies?  In the 3D
perovskite manganites such as $\rm Pr_{\it 1-x}Ca_{\it x}MnO_3$
\cite{jirak85,tomioka96}, the CE-type orbital and magnetic order
persist in a wide doping range ($0.3<x<0.7$). Doping this system
away from $x=0.5$ to form $\rm Pr_{0.7}Ca_{0.3}MnO_3$ induces a
ferromagnetic component below the AF CE-ordering temperature
\cite{yoshizawa98}. This has been interpreted as evidence for phase
coexistence involving ferromagnetic clusters and the AF CE structure
\cite{moreo99}, although the data are also consistent with a canted
antiferromagnet \cite{yoshizawa95,fernandezbaca02}.

In the case of the single-layer $\rm La_{\it 1-x}Sr_{\it 1+x}MnO_4$
(LSMO), extensive neutron and x-ray scattering experiments
\cite{sternlieb96,larochelle01,larochelle05,senff06,senff08} have
shown that the system at $x=0.5$ has a checkerboard charge
ordering-orbital ordering (CO-OO) and a CE-type spin configuration
similar to that of the perovskite manganites. However, as soon as
electrons are doped into the $x=0.5$ compound, the CE spin structure
is destroyed and the system behaves like a spin-glass while CO-OO is
maintained \cite{larochelle01}.  But is this a generic feature or
does it depend on the bandwidth and electron-phonon coupling of the
material?

In this Letter, we report a systematic neutron scattering study of
the magnetic correlations in the single-layer manganite $\rm Pr_{\it
1-x}Ca_{\it 1+x}MnO_4$ (PCMO). Contrary to LSMO, electron-doping
into the $x=0.5$ PCMO \cite{chi07,mathieu07,yu07} is found to
suppress, but not fully eliminate, the commensurate CE AF order. In
addition, a set of new incommensurate (ICM) magnetic peaks are
observed near the CM scattering from the CE structure. These results
are compatible with a self-organized mixed phase scenario where the
state at $x < 0.5$ is composed of CE patches separated by
electron-rich but magnetically disordered domain walls (Fig.~1).
This result establishes potentially interesting analogies between the
doping of some Cu oxide cuprates that leads to stripes, and the
doping of the CE state in manganites that also seems to induce
electron-rich domain walls.

\begin{figure}[ht!]
\includegraphics[width=3.0in]{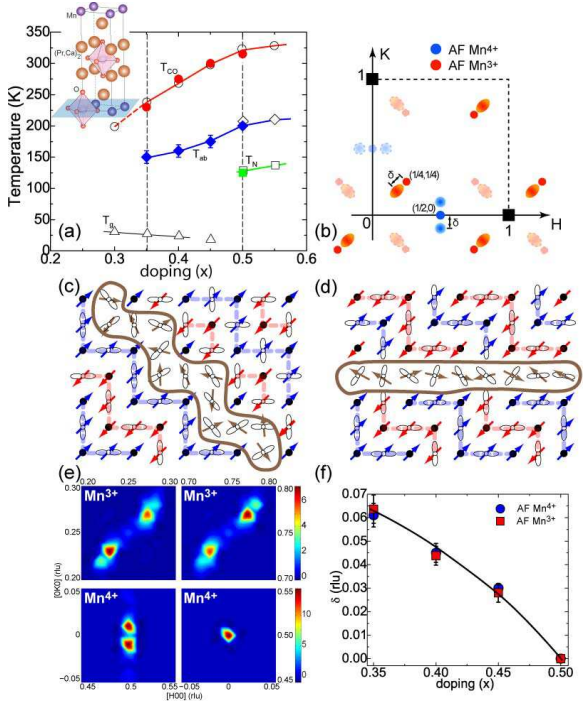}
\caption{\label{fig:fig1}
(a) Phase diagram of $\rm Pr_{\it 1-x}Ca_{\it 1+x}MnO_4$. Solid
symbols are from our measurements, open symbols from
Ref.~{[\onlinecite{mathieu07}]}. $\rm T_{CO}$ denotes the transition
temperature of CO-OO, $\rm T_{ab}$ and $\rm T_{N}$ are the
transition temperatures of 2D and 3D AF order. (b) Schematics of the
experimental observations in reciprocal space. The pattern in
lighter color originates from the 90-degree twinned domain. Orbital
and spin configurations within the $\rm MnO_2$-plane that may
explain the neutron results: (c) are diagonal and (d) horizontal
domain walls, where extra electrons congregate. Mn$^{4+}$ ions are
denoted by black circles. The Jahn-Teller active Mn$^{3+}$ ions have
additional orbitals. The blue and red lines illustrate the
ferromagnetic zig-zag spin chains which are coupled
antiferromagnetically. (e) Fourier transform of the spin
configurations in (c) and (d) combined. Note the dominant spectral
weight at $(1/4-\delta/\sqrt{2},1/4-\delta/\sqrt{2})$ or
$(3/4+\delta/\sqrt{2},3/4+\delta/\sqrt{2})$ for the Mn$^{3+}$ spins.
(f) Experimentally observed doping dependence of incommensurability
from both the Mn$^{3+}$ and Mn$^{4+}$ sublattices.} 
\end{figure}

Single crystals of PCMO (mass $\approx$ 4 to 6 grams) were grown
using the floating zone method.  The phase purity and cation
concentrations were checked by neutron, x-ray powder diffraction
and inductively coupled plasma atomic emission spectroscopy, no
noticable composition defects were found. At room temperature, PCMO
has an orthorhombic structure slightly distorted from the tetragonal
symmetry, with lattice parameters $a \approx 5.38$ \AA, $b \approx
5.40$ \AA, and $c \approx 11.85$ \AA.  For simplicity, we use the
tetragonal unit cell ($a=b\approx 3.84$ \AA).  The experiments were
carried out using triple-axis spectrometers at the NIST Center for
Neutron Research with final neutron energy fixed at $E_f=14.7$ or
13.5~meV.  The momentum transfers $q=(q_x,q_y,q_z)$ in units of
$\AA^{-1}$ are at positions
$(h,k,l)=(q_xa/2\pi,q_yb/2\pi,q_zc/2\pi)$ in reciprocal lattice
units (rlu). 

Figures 1(a)-1(b) summarize the main results of our neutron
scattering measurements. In $\rm Pr_{0.5}Ca_{1.5}MnO_4$, the Mn
spins form two magnetic sublattices. Compatible with a CE state, the
characteristic wavevector associated with the Mn$^{3+}$ spins
appears at $q_1=(1/4,/1,4,0)$ and the corresponding wavevector for
the Mn$^{4+}$ spins is at $q_2=(1/2,0,0)$. However, the electron
doping of $\rm Pr_{0.5}Ca_{1.5}MnO_4$ induces additional ICM spin
correlations at $(1/4-\delta/\sqrt{2},1/4-\delta/\sqrt{2},0)$ and
$(1/2,\pm\delta,0)$, respectively. The incommensurability (defined
as the distance $\delta$ between the CM and the ICM peaks) of both
sublattices shows a robust doping dependence [Figure 1(f)].

\begin{figure}[bh!]
\includegraphics[width=3.3in]{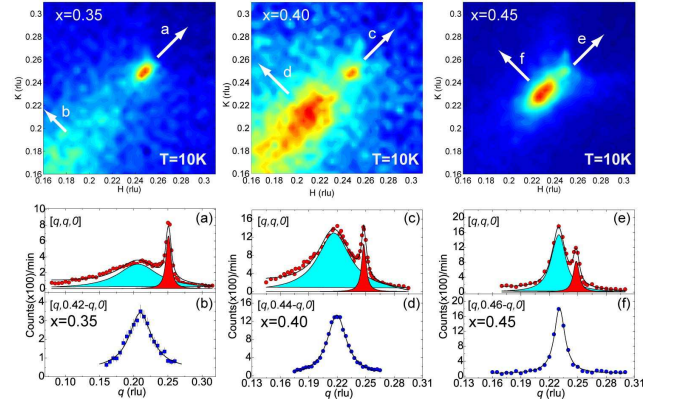}
\caption{\label{fig:fig2}
Comparison of the low-$T$  magnetic scattering originating from the
$\rm Mn^{3+}$ spins, for the $x=0.35$, $x=0.40$, and $x=0.45$
samples.  The corresponding wavevector scans through the peaks, as
indicated by the arrows, are presented in the lower panels. The data
reveal that a separate ICM component develops as electrons are added
from the $x=0.5$ point, which coexists with the $x=0.5$ AF order.
The ICM scattering is short range, and separates further from the CM
position and becomes broader as $x$ is reduced.  
} 
\end{figure}

Figure 2 shows the evolution of the magnetic scattering from the
Mn$^{3+}$ sublattice for the doped samples.  A wide range of
reciprocal space near the Bragg point (1/4,1/4,0) and equivalent
positions in higher Brillouin zones was surveyed. At $x=0.35$, the
additional scattering consists of two parts.  First, magnetic
scattering is located at positions identical to those observed at
$x=0.50$ \cite{chi07}, but with weaker magnetic correlations between
MnO$_2$ planes along the $c$-axis.  Second, there is ICM scattering
at wavevector $(1/4-\delta/\sqrt{2},1/4-\delta/\sqrt{2},0)$. This
scattering becomes strongest near (3/4,1/4,0) and decreases in
intensity near (3/4,3/4,0). The ICM scattering is highly anisotropic
with a rod-like profile elongated along the [1,1,0] (longitudinal)
direction indicating a shorter correlation length in this direction.
As the doping evolves toward $x$=0.5, the diffusive ICM scattering
sharpens and gradually moves towards the CM position. At $x=0.45$,
the scattering remains anisotropic, but the difference between the
two orthogonal directions becomes smaller. As displayed in the
wavevector scans, the longitudinal scan shows a broader width when
compared to the transverse scan along the [1,1,0] direction. The
corresponding correlation lengths $\xi_L$ and $\xi_T$ from the ICM
scattering, after deconvoluting the instrumental resolution, are
listed in Table~1. Both numbers increase substantially as the doping
approaches $x=0.5$.

\begin{table}[ht!]
\caption{Doping dependence of the magnetic scattering correlation
length $\xi$ from the Mn$^{3+}$ and Mn$^{4+}$ sublattices. ``$L$''
and ``$T$'' denote the longitudinal and transverse directions.
}
\label{tab1}
\begin{tabular}{lrrr}
\hline \hline
$\rm Pr_{\it 1-x}Ca_{\it 1+x}MnO_4$ & $x=0.35$    & $x=0.40$  & $x=0.45$  \\
  \hline
$\rm \xi_{L}(Mn^{3+})(\AA)$	 & $12.2\pm0.3$ & $21.1\pm0.4$ & $52.5\pm1.5$  \\
$\rm \xi_{T}(Mn^{3+})(\AA)$ 	 & $20.2\pm0.4$ & $37.6\pm0.9$ & $66.5\pm1.8$  \\
$\rm \xi_{CM}(Mn^{3+})(\AA)$ 	 & $116.8\pm4.5$ & $139.4\pm6.4$ & $127.1\pm6.6$ \\
\hline
$\rm \xi_{ICM}(Mn^{4+})(\AA)$ 	 & $9.8\pm1.1$  & $15.9\pm1.3$ & $44.6\pm1.4$ \\
$\rm \xi_{CM}(Mn^{4+})(\AA)$ 	 & $131.4\pm2.8$ & $127.3\pm5.3$ & $121.0\pm7.2$ \\
\hline \hline
\end{tabular}
\end{table}

\begin{figure}[bh!] 
\includegraphics[width=3.1in]{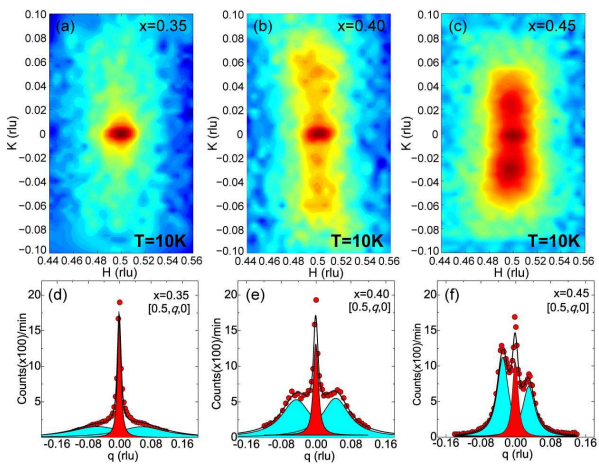}
\caption{\label{fig:fig3} 
Comparison of the magnetic diffuse scattering originating from the
$\rm Mn^{4+}$ spins at $T=10$~K for the (a) $x=0.35$, (b) $x=0.40$,
and (c) $x=0.45$ samples.  Wavevector scans across the ICM and
CM peaks are in the panels (d)-(f).
}
\end{figure}

The contour plots and the wavevector scans of the $\rm Mn^{4+}$
spins probed near $q_2=(1/2,0,0)$ are illustrated in Figure~3.  The
intense peaks at CM positions are surrounded by broad diffuse
scattering. The CM component is sharper than the ICM fluctuations,
but the Lorentzian profiles in the wavevector scans indicate the
lack of true long-range order. Similar to the Mn$^{3+}$ sublattice,
the incommensurability of the diffuse scattering can be tuned by
sample composition. As the ICM peaks move closer to the CM position
with increasing $x$, the width of the diffuse peak narrows and the
intensities become enhanced. The short-range correlations are less
anisotropic than those in the Mn$^{3+}$ sublattice.

\begin{figure}[ht!] \includegraphics[width=3.0in]{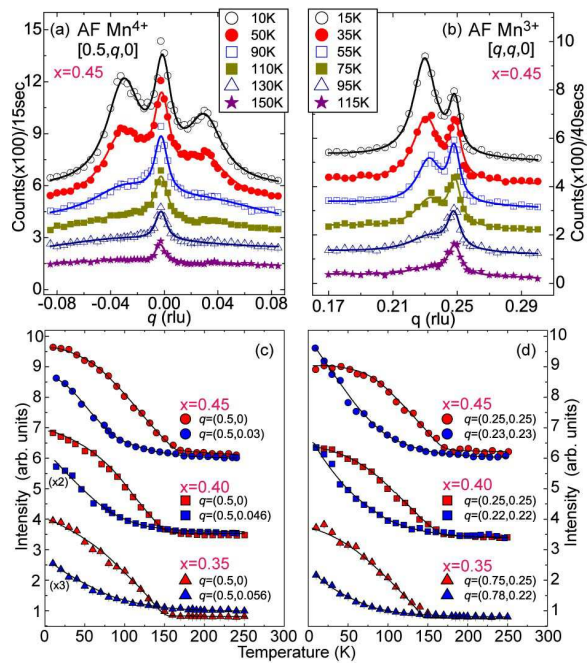}
\caption{\label{fig:model} 
Wavevector scans of magnetic diffuse scattering from (a) $\rm
Mn^{4+}$ and (b) $\rm Mn^{3+}$ sublattices of the $x=0.45$ sample at
selected temperatures. Temperature dependence of the peak
intensities of the CM (red) and ICM (blue) peaks from the (c) $\rm
Mn^{4+}$ and (d) $\rm Mn^{3+}$ sublattices for $x=0.35,0.40$, and
$0.45$.
} 
\end{figure}

To further characterize the magnetic correlations, we investigate
the $T$-dependence of the wavevector scans at $x=0.45$ in Figs.~4(a)
and 4(b). At $T=$10~K, distinct CM and ICM peaks are observed. Upon
warming, the amplitude of the ICM peak is rapidly suppressed with
little variation in peak position and the scattering evolves into a
broad feature as the temperature is raised.  The scattering profile
of the CM component, on the other hand, remains well resolved to a
much higher temperature.  The concave shape of the peak intensity
vs.~temperature from the ICM scattering [Figs.~4(c)-4(d)] reveals
the expected diffusive nature for the short-range correlations, in
contrast to the order-parameter-like CM scattering.  

The suppression of the long range CE magnetic order and the
surprising emergence of the ICM spin fluctuations in PCMO are
different from the insulating perovskite $\rm Pr_{\it 1-x}Ca_{\it
x}MnO_3$, where the CE-type spin order survives over a broad carrier
doping range \cite{tomioka96}. Our results suggest a form of
electronic self-organization in this single-layer compound.
One might speculate that the ICM scattering results from the
formation of an ICM charge density wave, with an associated spin
density wave (SDW) \cite{loudon05}. In this picture, the overall
magnetic structure would resemble the usual CE-type checkerboard
spin configuration, but the amplitudes of the spins would have a
smooth spatial modulation to accommodate the extra electrons.  While
such a configuration naturally brings about the ICM magnetic peaks,
they should be symmetric with respect to the CM positions \cite{sdw}
and the CE peaks should be absent, in contrast to the experimental
observations. An alternative model consists of a mixed phase of the
CE structure and the competing ferromagnetic state.  There is ample
evidence of this mixed state in the CMR regime near the Curie
temperature \cite{yoshizawa95,fernandezbaca02}. However, the present
case is for low temperatures, and more importantly, we do not
observe any ferromagnetic scattering. Therefore phase separation
involving large size CE and FM clusters does not provide an
explanation of our results either.

A model that is consistent with the overall experimental
observations involves the formation of a stripe phase similar to
those in the high-$T_{\rm c}$ superconducting cuprates
\cite{tranquada95}. For doping below $x=0.5$, the system would form
magnetic clusters or domains which preserve the CE-type spin
configuration with distinct neighboring Mn$^{3+}$ and Mn$^{4+}$
sites. The excess electrons would congregate at the domain
boundaries with random spin and orbital orientations. We explored a
variety of real-space spin arrangements and found two different ones
that, when combined, characterize the observations.  The first one
[Fig.~1(c)] describes a diagonal Mn$^{3+}$ domain boundary
separating two CE clusters in which the Mn$^{3+}$ spins are in
anti-phase while the Mn$^{4+}$ spins are not. The individual
magnetic domain contributes to the CM scattering, while the ICM peak
along the diagonal arises from the correlation between domains
\cite{dis}. This configuration reproduces the magnetic scattering
originating from the Mn$^{3+}$ sites, but leaves the scattering near
(1/2,0,0) undisturbed. The second one [Fig.~1(d)] describes spin
arrangements with a horizontal domain boundary. There is an extra
phase shift (1/4 or 3/4 of the 4$a_c$, the periodicity of the
CE-phase) between adjacent magnetic domains. It introduces two
symmetric ICM peaks at $q=(1/2,\pm\delta,0)$ but not at
$q=(0,1/2\pm\delta,0)$. As demonstrated in Fig.~1(e), the Fourier
transformation of the above described combined configurations
successfully creates a pattern qualitatively similar to the
experimental results \cite{incoherent}. Our identification of the
inhomogeneous and textured spin states, unveils another degree of
similarity with the cuprates in terms of electronic
self-organization and spin incommensurability \cite{mannella05}, and
provides a useful comparison to investigate the competing order and
complexity in the strongly correlated electron system.

It was pointed out that the quenched disorder is important in
determining the stability of the CE-type magnetic phase
\cite{alvarez06}. Monte Carlo simulations suggest that a small
amount of disorder or randomness in 2D or 3D systems may destroy
that phase. However, the inherent quenched disorder \cite{disorder}
in PCMO caused by A-site solid solution is small, ($1\sim 2
\times10^{-7}$ for the doping range we studied) because of their
comparable $\rm Pr^{3+}/Ca^{2+}$ ionic size.  Therefore, the
preservation of CE-type fluctuations at lower doping in PCMO
confirms that the CE-phase could be a robust feature, in contrast to
its quick disappearance in LSMO with more quenched disorder
\cite{larochelle01}. On the other hand, the striking difference
between single-layer PCMO and the 3D perovskite manganites
highlights the crucial role of the magnetic interactions between
planes, which is believed to stabilize the CE-type order
\cite{senff06,ma09}.

\begin{acknowledgments}
We are grateful to D. Khomskii, Y. Ren, and M. Braden for their
helpful discussions. The experimental work was partially supported
by the Division of Scientific User Facilities of the Office of Basic
Energy Sciences, US Department of Energy and by the U.S. NSF
DMR-0756568 and DOE Nos.  DE-FG02-05ER46202 grants. The theory
effort was supported by the NSF grant DMR-0706020, and by the Div.
of Materials Sciences and Eng., U.S. DOE under contract with
UT-Battelle, LLC. 
\end{acknowledgments}


\begin{references}
\bibitem{sdw} An homogeneous incommensurate spin configuration for
the Mn$^{4+}$ ions in the CE pattern that produces peaks in the
structure factor at $(1/2,\pm\delta,0)$ as in Fig.~3 arises from
correlation functions that in real space have the form
$S(x,y,0)=A(-1)^x cos(\delta y)$. However, a Fourier transform of
this expression produces a structure factor with peaks at
$(1/2,\pm\delta,0)$ and also at $(0,1/2\pm\delta,0)$ which is not
observed in the experiments.
\bibitem{dis} A small amount of disorder to the spin orientations
about the CE state has been added to make the system more realistic
but similar results are obtained without disorder.
\bibitem{incoherent} Note that our model combines two configurations
``incoherently''. A more complicate ``coherent'' model might reveal
other novel features (like extra satellite peaks) and deserves future
investigation.
\bibitem{disorder} This is characterized by $ \sigma^2=\sum_i (x_i
r_i^2 - {\bar r}^2)$, where $x_i$ is the fractional occupancies of
$A$-site species, $r_i$ and $\bar r=\sum_i x_i r_i$ are individual
radii and averaged ionic radius.
\end{references}
\end{document}